\documentclass[conference]{IEEEtran}
\IEEEoverridecommandlockouts
\def\BibTeX{{\rm B\kern-.05em{\sc i\kern-.025em b}\kern-.08em
    T\kern-.1667em\lower.7ex\hbox{E}\kern-.125emX}}
\usepackage{amsmath, amsfonts, amssymb, xcolor, amsthm, balance, mathrsfs, multirow, lipsum, cite, graphicx} 
\let \bs \boldsymbol
\usepackage[hyphens]{url}  
\usepackage{concmath}
\usepackage[T1]{fontenc} 
\begin{document}

%\title{Conference Paper Title*\\
%{\footnotesize \textsuperscript{*}Note: Sub-titles are not captured in Xplore and
%should not be used}
%\thanks{Identify applicable funding agency here. If none, delete this.}
%}

\title{Link-Layer Capacity of Downlink NOMA with Generalized Selection Combining Receivers\thanks{This publication has emanated from research conducted with the financial support of Science Foundation Ireland (SFI) and is co-funded under the European Regional Development Fund under Grant Number 13/RC/2077.}}

\author{\IEEEauthorblockN{Vaibhav Kumar\IEEEauthorrefmark{1}, Barry Cardiff\IEEEauthorrefmark{1}, Shankar Prakriya\IEEEauthorrefmark{2}, and Mark F. Flanagan\IEEEauthorrefmark{1}}
\IEEEauthorblockA{\IEEEauthorrefmark{1}School of Electrical and Electronic Engineering, 
University College Dublin, Belfield, Dublin 4, Ireland\\
\IEEEauthorrefmark{2}Department of Electrical Engineering, Indian Institute of Technology -- Delhi, New Delhi, India\\
Email: vaibhav.kumar@ucdconnect.ie, barry.cardiff@ucd.ie, shankar@ee.iitd.ernet.in, mark.flanagan@ieee.org}}

\maketitle

\begin{abstract}
Non-orthogonal multiple access (NOMA) has drawn tremendous attention, being a potential candidate for the spectrum access technology for the fifth-generation (5G) and beyond 5G (B5G) wireless communications standards. Most research related to NOMA focuses on the system performance from Shannon's capacity perspective, which, although a critical system design criterion, fails to quantify the effect of delay constraints imposed by future wireless applications. In this paper, we analyze the performance of a single-input multiple-output (SIMO) two-user downlink NOMA system, in terms of the link-layer achievable rate, known as effective capacity (EC), which captures the performance of the system under a delay-limited quality-of-service (QoS) constraint. For signal combining at the receiver side, we use generalized selection combining (GSC), which bridges the performance gap between the two conventional diversity combining schemes, namely selection combining (SC) and maximal-ratio combining (MRC). We also derive two approximate expressions for the EC of NOMA-GSC which are accurate at low-SNR and at high-SNR, respectively. The analysis reveals a tradeoff between the number of implemented receiver radio-frequency (RF) chains and the achieved performance, and can be used to determine the appropriate number of paths to combine in a practical receiver design.
\end{abstract}


\section{Introduction}
Worldwide efforts to enable 5G and B5G wireless communications are well underway, including new spectrum allocation policies, flexible modulation techniques, multiple-access  schemes, standardization, and implementation. NOMA has received tremendous attention as a design paradigm for the radio access techniques of future communications standards due to its inherent potential for higher spectral efficiency, denser connectivity and lower latency~\cite{ProcHanzo}. In contrast to the traditional \emph{orthogonal} radio-access schemes, multiple users in NOMA are served in the same resource block (time slot, subcarrier, frequency band, or spreading code). It was shown in~\cite{NOMA_book} that when the channels from the source to each of the multiple users are significantly different, NOMA can achieve a higher spectral efficiency as compared to its orthogonal multiple access (OMA) counterpart. In order to further enhance the spectral efficiency, different advanced signal processing schemes have been suggested, including cooperative NOMA~\cite{Coop_NOMA}, multi-antenna-assisted architectures for NOMA~\cite{MenGe} and multiple-input multiple-output NOMA (MIMO-NOMA)~\cite{MIMO-NOMA}. Most existing research related to the performance analysis of NOMA systems has dealt primarily with the metrics of outage probability, achievable rate and system throughput.

Although the classical (Shannon) ergodic capacity has been extremely successful in communication system design, it fails to explicitly characterize the delay-constrained performance, which is one of the most crucial system design parameters for 5G and B5G networks. To facilitate the investigation of wireless networks under statistical QoS limitations (e.g., data rate, delay and delay-violation probability), a link-layer analysis tool, known as \emph{effective capacity}, was first introduced in~\cite{Negi}. Given a delay-violation probability, the EC defines the maximum data arrival rate that can be supported by a radio link. The EC analysis of a multiple-input single-output (MISO) system in Nakagami-$m$, Rician and generalized-$\mathcal K$ fading was presented in~\cite{Larsson}. Recently, the EC analysis of a MISO system under the Fisher-Snedecor $\mathcal F$ fading was presented in~\cite{Snedecor}.
%The EC analysis of multiple-antenna Gaussian channels was presented in~\cite{Gaussian}. The EC analysis of an interference-and-delay constrained cognitive radio relay channel under both peak and average interference constraints was presented in~\cite{CR}. An analytical framework for the EC analysis of a multiple-input single-output (MISO) system under different fading models (including Nakagami-$m$, Rician and Generalized-$\mathcal K$) was proposed in~\cite{Larsson}. A perforamnce analysis of equal-gain combining (EGC) under generalized-Gamma, $\alpha$-$\kappa$-$\mu$ and $\alpha$-$\eta$-$\mu$ fading channels in terms of EC was presented in~\cite{VTC}. Recently, the EC analysis of MIMO channels with arbitrary inputs, and MISO systems over Fisher-Snedecor $\mathcal F$ fading channels were presented in~\cite{MIMO_EC} and~\cite{Snedecor}, respectively.

The available literature on the performance analysis of NOMA in terms of EC is relatively limited. In~\cite{NOMA_EC_Mussavian}, the achievable link-layer rate of a multi-user downlink NOMA network under a per-user delay QoS requirement was presented, where the users were divided into multiple NOMA pairs and OMA was applied for multiple access among each such NOMA pair. It was shown that in such a scenario, OMA outperforms NOMA in terms of the total link-layer rate in the low signal-to-noise-ratio (SNR) regime, while at high SNR, the NOMA system achieves higher EC as compared to its OMA-based counterpart. It was shown in~\cite{SubOptimal} that for a two-user downlink NOMA system, the optimal power control problem to maximize the sum effective capacity is non-convex and thus the notion of \emph{partial effective capacity} was introduced to derive a sub-optimal power allocation policy. For the case of half/full-duplex two-user cooperative NOMA, in order to maximize the minimum effective capacity of the user-pair, a bisection-based optimal power allocation scheme was proposed in~\cite{OptimalPower}. Analysis of the delay-violation probability and EC for a multi-user downlink NOMA system using stochastic network calculus (SNC) was presented in~\cite{NOMA_JSTSP}. The analysis of \emph{effective secrecy rate} for a multi-user downlink NOMA system was recently presented in~\cite{NOMA_Secrecy}.

In all of the aforementioned research related to the EC analysis of NOMA, a single-input single-output (SISO) system model was considered. In this paper, we consider a single-input multiple-output (SIMO) downlink NOMA system consisting of two users\footnote{Note that a two-user downlink version of NOMA, called \emph{multiuser superposition transmission} (MUST), has been proposed for the Third Generation Partnership Project Long Term Evolution Advanced (3GPP-LTE-A) standard.}. Although MRC is an optimal receiver combining technique for a SIMO system, it is highly susceptible to channel estimation error for the paths having lower instantaneous received SNR. On the other hand, SC uses the best path (in terms of SNR) and hence fails to exploit the total diversity offered by the other independent paths. A comparatively more flexible diversity combining scheme, called GSC, bridges this gap between SC and MRC, by adaptively combining a subset of the strongest paths. Therefore, GSC receivers are more robust to channel estimation errors and also require fewer active RF chains at the receiver, thereby reducing the overall cost of the system implementation. A detailed moment generating function (MGF)-based performance analysis of GSC in Rayleigh fading channels, in terms of bit error rate (BER), symbol error rate (SER) and outage probability, was presented in~\cite{MGF-GSC}. The performance analysis of GSC in a downlink NOMA system in terms of average achievable sum-rate and outage probability was presented in~\cite{EL}.  

Against this background, the main contributions in this paper are listed below:
\begin{itemize}
	\item We analyze the sum effective capacity of a two-user donwlink NOMA system with GSC receivers. For the case of the symbol intended for the strong user, we derive an exact closed-form expression for the EC. On the other hand, for the case of the symbol intended for the weak user, since it is somewhat tedious to find an exact closed-form expression for the EC in NOMA-GSC, we derive closed-form expressions for two special cases, namely NOMA-SC and NOMA-MRC. We also show that most of the gain (in terms of sum EC) is achieved by combining the strongest diversity paths, and that diminishing returns are obtained as the number of combined paths increases.
	\item We derive low-SNR as well as high-SNR approximations for the sum EC of NOMA-GSC. From the analysis of the EC for the symbol intended for the weak user, it can be noted that no significant benefit of multiple receive antennas is observed at the weak user. Also, the results indicate indicate that the sum EC grows exponentially with SNR in the low-SNR regime, and linearly with SNR in the high-SNR regime.
	\item Using Jensen’s inequality, we derive a fundamental upper-bound on the sum EC of NOMA-GSC which is independent of the delay constraint. We show that this bound is equal to the average achievable sum rate of NOMA-GSC and that the difference between the achievable sum rate and the sum EC increases with increasing SNR and with an increase in the delay exponent.
	\item For the purpose of comparison, we also derive a closed-form expression for the sum EC of a two-user downlink OMA-GSC system, and show that NOMA-GSC outperforms OMA-GSC in terms of sum EC for any number of combined paths. We show that as the delay constraint becomes more strict, the performance difference between NOMA-GSC and OMA-GSC becomes smaller.
\end{itemize}
%%=============System Model===========
\section{System Model}
Consider the two-user downlink NOMA system consisting of a source $S$ equipped with a single transmit antenna, and two users $U_s$ (the user with  \emph{strong} channel conditions) and $U_w$ (the user with \emph{weak} channel conditions) equipped with $N_s \geq 1$ and $N_w \geq 1$ receive antennas, respectively. We assume a half-duplex communication protocol. The channel coefficients between $S$ and $U_s$ are assumed to be independent and identically distributed (i.i.d.) according to $\mathcal{CN}(0, \Omega_s)$ with mean-square value $\Omega_s$, while those between $S$ and $U_w$ are assumed to be i.i.d. according to $\mathcal{CN}(0, \Omega_w)$ with mean-square value equal to $\Omega_w$. We assume that $\Omega_w < \Omega_s$, and that for each link perfect CSI is available at the receiver side, whereas only the knowledge of $\Omega_s$ and $\Omega_w$ is available at the transmitter side\footnote{This is different from the two-user NOMA system presented in~\cite{NOMA_EC_Mussavian}, where (perfect) \emph{instantaneous} CSI was assumed to be available at the transmitter and therefore the users were ordered instantaneously (in terms of strong and weak users). In our system, the users are ordered depending on the \emph{statistical} CSI. This reduces the signaling overhead at the transmitter and the overall system implementation cost.}.  

The source transmits a power-domain multiplexed symbol $\sqrt{a_s \mathcal E} x_s + \sqrt{a_w \mathcal E} x_w$ to both users, where $x_i, i \in \{s, w\}$ denotes the information-bearing constellation symbol intended for user $U_i$ (here we assume that $\mathbb E\{|x_i|^2\} = 1$), $\mathcal E$ is the energy budget at the source for each time slot, $a_i$ is the power allocation coefficient with $a_s < a_w$ and $\sum_i a_i = 1$. Upon signal reception, user $U_s$ (resp. $U_w$) adaptively selects a subset of $n_s, 1 \leq n_s \leq N_s$ (resp. $n_w, 1 \leq n_w \leq N_w$) antennas, out of the total $N_s$ (resp. $N_w$) antennas which provide the strongest links (in terms of the received SNR), and then applies MRC to combine the received signals. Therefore, the signals received at user $U_i$ (after applying GSC) is given by 
\begin{align*}
	\bs y_i = \bs h_{i}^H \left[ \bs h_i \left( \sqrt{a_s \mathcal E} x_s + \sqrt{a_w \mathcal E} x_w \right) + \bs z_i \right],
\end{align*}
where $\bs h_i = [h_{i, 1} \, h_{i, 2} \, \ldots, \, h_{i, n_i}]^T \in \mathbb C^{n_i \times 1}$ contains the $n_i$ largest-magnitude channel coefficients between $S$ and $U_i$ in \emph{decreasing} order, i.e., $|h_{i, 1}| \geq |h_{i, 2}| \geq \cdots \geq |h_{i, n_i}|$, and $\bs z_i = [z_{i, 1} \, z_{i, 2} \, \ldots, \, z_{i, n_i}]^T \in \mathbb C^{n_i \times 1}$. Each element in $\bs z_i$ represents additive white Gaussian noise (AWGN) with zero mean and variance $\sigma^2$.

Both $U_s$ and $U_w$ first decode $x_w$ considering the interference from $x_s$ as additional noise. User $U_s$ then applies successive interference cancellation (SIC) to decode the intended symbol $x_s$. Since the symbol $x_w$ needs to be decoded correctly at both users (at $U_w$ as the intended symbol and at $U_s$ for SIC), while $x_s$ needs to be correctly decoded only at user $U_s$, the received instantaneous SINR and SNR for the correct decoding of $x_w$ and $x_s$ are, respectively, given by
\begin{align}
	\gamma_w = \dfrac{a_w g_{\min} \rho}{a_s g_{\min} \rho + 1}, \qquad \gamma_s = a_s g_s \rho, \notag 
\end{align}
where $g_i \triangleq \bs h_i^H \bs h_i$, $g_{\min} \triangleq \min\{g_s, g_w\}$ and $\rho \triangleq \mathcal E/\sigma^2$. Using a transformation of random variables and~\cite[eqn.~(16)]{MGF-GSC}, the probability density function (PDF) of $g_i$ can be given by
\begin{align}
	& f_{g_i}(x) = \binom{N_i}{n_i} \left[ \dfrac{x^{n_i - 1} \exp \left( \tfrac{-x}{\Omega_i}\right)}{\Omega_i^{n_i} (n_i - 1)!}  + \dfrac{1}{\Omega_i} \sum_{l = 1}^{N_i - n_i} (-1)^{n_i +l - 1} \right. \notag \\
	& \times \binom{N_s - n_s}{l}\left( \dfrac{n_i}{l}\right)^{n_i - 1} \exp \left( \dfrac{-x}{\Omega_i} \right) \left\{ \exp \left( \dfrac{-l x}{n_i \Omega_i}\right) \right. \notag \\
	& \hspace{4cm}\left. \left. - \sum_{m = 0}^{n_i - 2}  \dfrac{1}{m!} \left( \dfrac{-l x}{n_i \Omega_i} \right)^m\right\} \right]. \label{f_gi}
\end{align}
%\begin{figure*}[t]
%\normalsize
%\setcounter{MYtempeqncnt}{\value{equation}}
%\begin{align}
%	f_{g_i}(x) = & \ \binom{N_i}{n_i} \left[ \dfrac{x^{n_i - 1} \exp \left( \tfrac{-x}{\Omega_i}\right)}{\Omega_i^{n_i} (n_i - 1)!}  + \dfrac{1}{\Omega_i} \sum_{l = 1}^{N_i - n_i} (-1)^{n_i +l - 1} \left( \dfrac{n_i}{l}\right)^{n_i - 1} \exp \left( \dfrac{-x}{\Omega_i} \right) \left\{ \exp \left( \dfrac{-l x}{n_i \Omega_i}\right) - \sum_{m = 0}^{n_i - 2}  \dfrac{1}{m!} \left( \dfrac{-l x}{n_i \Omega_i} \right)^m\right\} \right]. \label{f_gi}
%\end{align}
%\setcounter{equation}{\value{MYtempeqncnt}}
%\hrulefill
%\end{figure*}
%\addtocounter{equation}{1} 

%=============Performance analysis===================
\section{Performance Analysis}
In this section, we present a comprehensive performance analysis of the two-user NOMA-GSC system in terms of effective capacity. We also derive closed-form approximations to the sum EC of NOMA-GSC, which are valid in the low-SNR and high-SNR regimes respectively, as well as an upper-bound on the sum EC. For a fair comparison, we also present the sum EC analysis of an OMA-GSC system.
%------------------------------
\subsection{Effective capacity of NOMA-GSC}
The expression for EC of symbol $x_i$ in NOMA-GSC can be given by~(c.f.~\cite{NOMA_EC_Mussavian})
\begin{align}
	E_i = -\dfrac{1}{\theta T B} \ln \left[ \mathbb E \left\{ \exp \left( -\theta T B R_i\right) \right\} \right], \label{Ei_Def_Basic}
\end{align}
where $\theta$ is the delay QoS exponent (denotes the asymptotic delay-rate of the buffer occupancy at the transmitter, defined as $\theta \triangleq -\lim_{x \to \infty} \ln \Pr\{L > x\}/x$, $L$ being the equilibrium queue-length of the buffer at the transmitter), $T$ is the length of each fading-block (this is assumed to be same for all the wireless links and an integer multiple of the symbol duration\footnote{We also assume that the symbol durations for both $x_s$ and $x_w$ are the same.}), $B$ is the total available bandwidth, and $R_i$ is the instantaneous achievable rate for symbol $x_i$ (defined as $R_i = \log_2\{1 + \gamma_i\}$). It is important to note that $\theta \to \infty$ represents a system with very stringent delay constraint, while $\theta \to 0$ corresponds to the system with no delay constraint. For the case when $\theta \to 0$, the effective capacity becomes equal to the average achievable rate. Substituting the expression for the instantaneous achievable rate into~\eqref{Ei_Def_Basic}, the expression for the effective capacity of symbol $x_i$ in NOMA-GSC can be given by 
\begin{align}
	E_i = -\dfrac{1}{\nu} \log_2\left[ \mathbb E_{\gamma_i} \left\{ \left( 1 + \gamma_i \right)^{-\nu} \right\}\right], \label{Ei_Def}
\end{align}
where $\nu \triangleq \theta T B/\ln 2$. Therefore, for $x_s$, we have 
\begin{align}
	E_s = -\dfrac{1}{\nu} \log_2 \left[ \int_0^\infty (1 + x)^{-\nu} f_{\gamma_s}(x) \mathrm dx \right]. \label{Es_Def}
\end{align}
Substituting the expression for $f_{g_s}(x)$ from~\eqref{f_gi} into~\eqref{Es_Def}, a closed-form expression for $E_s$ can be given by
\begin{align}
	& \ E_{s} \!=\! \dfrac{-1}{\nu} \log_2 \left[ \! \binom{N_s}{n_s} \! \left\{ \dfrac{I_1}{\Omega_s^{n_s} (n_s - 1)!} \! + \! \dfrac{1}{\Omega_s} \!\! \sum_{l = 1}^{N_s - n_s} \! \!\! (-1)^{n_s + l - 1} \right. \right. \notag \\
	& \left. \left. \times \binom{N_s \!-\! n_s}{l} \! \left( \dfrac{n_s}{l} \right)^{n_s - 1} \!\! \left\{ \! I_2 (l) - \!\!\! \sum_{m = 0}^{n_s - 2} \! \dfrac{I_3(m)}{m!} \left( \dfrac{-l}{n_s \Omega_s} \right)^{\!\!m}  \right\} \right\} \right], \label{Es_Closed}
\end{align}
where $I_1 \triangleq \int_0^\infty (1 + a_s \rho x)^{-\nu} \ x^{n_s - 1} \exp \left( \tfrac{-x}{\Omega_s} \right) \mathrm dx = \tfrac{1}{\Gamma(\nu)}\int_0^\infty x^{n_s - 1} G_{1, 1}^{1, 1} \left( a_s \rho x \left\vert \begin{smallmatrix}1-\nu \\ 0 \end{smallmatrix}\right.\right) G_{0, 1}^{1, 0} \left( \tfrac{x}{\Omega_s} \left\vert \begin{smallmatrix} - \\[0.6em] 0 \end{smallmatrix}\right. \right) \mathrm dx = \tfrac{1}{\Gamma(\nu) (a_s \rho)^{n_s}} G_{1, 2}^{2, 1} \left( \tfrac{1}{\Omega_s a_s \rho} \left\vert \begin{smallmatrix} 1 - n_s \\ 0, \ \nu - n_s  \end{smallmatrix} \right.\right)$, $I_2 (l) \triangleq \int_0^\infty (1 + a_s \rho x)^{-\nu} \exp \left[ \tfrac{-1}{\Omega_s} \left( 1 + \tfrac{l}{n_s} \right)x \right] \mathrm dx =  \tfrac{\Phi_{s,l}^{\nu - 1}}{(a_s \rho)^\nu} \exp \left( \Phi_{s,l}/a_s \rho \right) \Gamma \left( 1- \nu, \Phi_{s,l}/a_s \rho \right)$ and $I_3 (m) \triangleq  \int_0^\infty (1 + a_s \rho x)^{-\nu} \exp \left( -x/\Omega_s \right) x^m \ \mathrm dx = \tfrac{1}{\Gamma(\nu) (a_s \rho)^{m + 1}} G_{1, 2}^{2, 1} \left( \tfrac{1}{\Omega_s a_s \rho} \left\vert \begin{smallmatrix} -m \\ 0, \ \nu - m - 1  \end{smallmatrix} \right.\right)$. Here $G_{\cdot, \cdot}^{\cdot, \cdot}(\cdot)$ denotes Meijer's G-function, $\Gamma(\cdot, \cdot)$ represents the upper-incomplete Gamma function and $\Phi_{i, l} \triangleq (1 + \tfrac{l}{n_i})/\Omega_i$. The integration in $I_1$ is solved using~\cite[eqns.~(7),~(10),~(11),~(21) and~(22)]{Reduce}, the integration in $I_2(l)$ is solved using~\cite[eqn.~(3.382-4),~p.~347]{Grad}, and the integration in $I_3(m)$ is solved similarly to $I_1$. 

On the other hand, using~\eqref{Ei_Def}, the EC for $x_w$ can be given by 
\begin{align}
	\!\!\!\!E_w = -\dfrac{1}{\nu} \log_2 \left[ \int_0^\infty \left( 1 + \dfrac{a_w \rho x}{a_s \rho x + 1} \right)^{-\nu} f_{g_{\min}}(x) \mathrm dx \right]. \label{Ew_Def}
\end{align}
The PDF of $g_{\min}$ can be given by $f_{g_{\min}}(x) = f_{g_w}(x)[1 - F_{g_s}(x)] + f_{g_s}(x)[1 - F_{g_w}(x)]$. It can be noted that the expression for the PDF of $g_{\min}$ is very complicated and hence deriving an exact closed-form expression for $E_w$ is somewhat tedious. Therefore, we discuss two special cases for the EC of $x_w$. 
\paragraph*{Case I ($n_s = n_w = n = 1$)} In this case, the NOMA-GSC system reduces to the NOMA-SC system where the signal from a single path (which has the highest instantaneous received SINR) is selected at both $U_s$ and $U_w$. Defining $g_{\min, \mathrm{SC}} \triangleq \min \{|h_{s, 1}|^2, |h_{w, 1}|^2\}$, using a transformation of random variables, it can be shown that $f_{g_{\min, \mathrm{SC}}}(x) = \sum_{k = 1}^{N_s} \sum_{j = 1}^{N_w} (-1)^{k + j} \binom{N_s}{k} \binom{N_w}{j} \chi_{k, j} \exp(-\chi_{k, j} x)$, where $\chi_{k, j} \triangleq \tfrac{k}{\Omega_s} + \tfrac{j}{\Omega_w}$. Therefore, replacing $f_{g_{\min}}(x)$ in~\eqref{Ew_Def} by $f_{g_{\min, \mathrm{SC}}}(x)$, a closed-form expression for the EC of $x_w$ in a NOMA-GSC system with $n = 1$ can be given by 
\begin{align*}
	& E_{w, \mathrm{SC}} = \dfrac{-1}{\nu} \log_2 \left[ \dfrac{1}{\Gamma(\nu) \Gamma(-\nu)} \sum_{k = 1}^{N_s} \sum_{j = 1}^{N_w} (-1)^{k + j} \binom{N_s}{k}  \right. \\
	& \hspace{0.5cm}\left. \times \ \binom{N_w}{j} \dfrac{\chi_{k, j}}{ \rho} \ \mathcal G_{1, 1:1, 1:0, 1}^{1, 1:1, 1:1, 0} \left( \begin{smallmatrix} 0 \\[0.6em] \nu - 1 \end{smallmatrix} \left\vert  \begin{smallmatrix} 1 + \nu \\[0.6em] 0  \end{smallmatrix} \right\vert \left. \begin{smallmatrix} - \\[0.6em] 0\end{smallmatrix}\right\vert a_s, \dfrac{\chi_{k, j}}{\rho} \right) \right],
\end{align*}
where $\mathcal G_{\cdot, \cdot:\cdot, \cdot:\cdot, \cdot}^{\cdot, \cdot:\cdot, \cdot:\cdot, \cdot}(\cdot)$ denotes the extended generalized bivariate Meijer G-function (EGBMGF) and the integral is solved using~\cite[eqns.~(10),~(11)]{Reduce} and~\cite[eqn.~(9)]{KappaMuShadowed}.
\paragraph*{Case II ($n_s = N_s$, $n_w = N_w$)} In this case, the NOMA-GSC system reduces to the NOMA-MRC system where signals from all the diversity paths are combined. Defining $g_{\min, \mathrm{MRC}} \triangleq \min_{i \in \{s, w\}} \left\{ \sum_{n = 1}^{N_i} |h_{i, n}|^2\right\}$, using a transformation of random variables it can be shown that $f_{g_{\min, \mathrm{MRC}}}(x) = \tfrac{x^{N_s - 1}}{\Gamma(N_s) \Omega_s^{N_s}} \exp (-\chi_{1, 1} x) \sum_{j = 0}^{N_w - 1} \tfrac{x^j}{j! \Omega_w^j} + \tfrac{x^{N_w - 1}}{\Gamma(N_w) \Omega_w^{N_w}} \exp (-\chi_{1, 1} x) \sum_{k = 0}^{N_s - 1} \tfrac{x^k}{k! \Omega_s^k}$. Therefore, replacing $f_{g_{\min}}(x)$ in~\eqref{Ew_Def} by $f_{g_{\min, \mathrm{MRC}}}(x)$ , a closed-form expression for the EC of $x_w$ in a NOMA-GSC system with $n_s = N_s$ and $n_w = N_w$ can be given by 
\begin{align*}
	& \ E_{w, \mathrm{MRC}} = \dfrac{-1}{\nu} \log_2 \left[ \dfrac{1}{\Gamma(N_s) \Omega_s^{N_s}} \sum_{j = 0}^{N_w - 1} \dfrac{\rho^{-(N_s + j)}}{j! \Omega_w^j}  \right. \\
	& \times \mathcal G_{1, 1:1, 1:0, 1}^{1, 1:1, 1:1, 0} \left( \begin{smallmatrix} 1 - (N_s + j) \\[0.6em] \nu - (N_s + j) \end{smallmatrix} \left\vert  \begin{smallmatrix} 1 + \nu \\[0.6em] 0  \end{smallmatrix} \right\vert \left. \begin{smallmatrix} - \\[0.6em] 0\end{smallmatrix}\right\vert a_s, \dfrac{\chi_{k, j}}{\rho} \right) \! + \dfrac{1}{\Gamma(N_w) \Omega_w^{N_w}} \\
	& \!\! \!\! \times \sum_{k = 0}^{N_s - 1} \!\! \dfrac{\rho^{-(N_w + k)}}{k! \Omega_s^k}   \left.  \mathcal G_{1, 1:1, 1:0, 1}^{1, 1:1, 1:1, 0} \left( \!\! \begin{smallmatrix} 1 - (N_w + k) \\[0.6em] \nu - (N_w + k) \end{smallmatrix} \! \left\vert  \!\begin{smallmatrix} 1 + \nu \\[0.6em] 0  \end{smallmatrix} \right\vert \! \left. \begin{smallmatrix} - \\[0.6em] 0\end{smallmatrix}\right\vert a_s, \dfrac{\chi_{k, j}}{\rho} \right) \right].	
\end{align*}

The sum effective capacity for NOMA-GSC is then obtained as $E_{\mathrm{sum}} = E_s + E_w$.
%-----------------------------------------------------
\subsection{Effective capacity of OMA-GSC}
For a fair comparison, we also present a closed-form analysis for the EC of OMA-GSC in this subsection. Considering a time-division multiplexed system, $S$ transmits $\sqrt{\mathcal E} x_w$ and $\sqrt{\mathcal E} x_s$ in the first and second time slots, respectively. The instantaneous achievable rate for $x_i$ is given by $\hat R_i = 0.5 \log_2(1 + \hat \gamma_i)$, where $\hat \gamma_i = \rho g_i$. Therefore, similar to~\eqref{Ei_Def}, the EC of $x_i$ for OMA-GSC can be given by 
\begin{align}
	\hat E_i = & \ - \dfrac{1}{\nu} \log_2 \left[ \mathbb E_{\hat \gamma_i} \left\{ (1 + \hat \gamma_i)^{-\nu/2} \right\} \right] \notag \\
	= & \ - \dfrac{1}{\nu} \log_2 \left[\int_0^\infty (1 + \rho x)^{-\nu/2} f_{g_i}(x) \mathrm dx\right]. \label{Ei_hat}
\end{align}
Substituting the expression for $f_{g_i}(x)$ from~\eqref{f_gi} into~\eqref{Ei_hat}, a closed-form expression for $\hat E_i$ is given by
\begin{align}
	& \hat E_{i} = \!  \dfrac{-1}{\nu} \log_2 \left[ \! \binom{N_i}{n_i} \! \left\{ \dfrac{\hat I_{i, 1}}{\Omega_i^{n_i} (n_i - 1)!} \! + \! \dfrac{1}{\Omega_i} \!\! \sum_{l = 1}^{N_i - n_i} \! \!\! (-1)^{n_i + l - 1} \right. \right. \notag \\
	& \times \!\! \! \left. \left. \binom{\!N_i \!-\! n_i\!}{l} \! \left( \dfrac{n_i}{l} \right)^{n_i - 1} \!\! \left\{ \! \hat I_{i, 2} (l) - \!\!\! \sum_{m = 0}^{n_i - 2} \! \dfrac{\hat I_{i,3}(m)}{m!} \left(\! \dfrac{-l}{n_i \Omega_i} \!\right)^{\!\!m}  \right\} \right\} \right], \label{Ei_hat_Closed}
\end{align}
%\begin{figure*}[t]
%\normalsize
%\setcounter{MYtempeqncnt}{\value{equation}}
%\begin{align}
%	\hat E_{i} = \!  \dfrac{-1}{\nu} \log_2 \left[ \! \binom{N_i}{n_i} \! \left\{ \dfrac{\hat I_{i, 1}}{\Omega_i^{n_i} (n_i - 1)!} \! + \! \dfrac{1}{\Omega_i} \!\! \sum_{l = 1}^{N_i - n_i} \! \!\! (-1)^{n_i + l - 1} \binom{N_i \!-\! n_i}{l} \! \left( \dfrac{n_i}{l} \right)^{n_i - 1} \!\! \left\{ \! \hat I_{i, 2} (l) - \!\!\! \sum_{m = 0}^{n_i - 2} \! \dfrac{1}{m!} \left( \dfrac{-l}{n_i \Omega_i} \right)^{\!\!m} \hat I_{i,3}(m) \right\} \right\} \right]. \label{Ei_hat_Closed}
%\end{align}
%\setcounter{equation}{\value{MYtempeqncnt}}
%\hrulefill
%\end{figure*}
%\addtocounter{equation}{1} 
where $\hat I_{i,1} \triangleq \int_0^\infty (1 + \rho g_i)^{-\nu/2} x^{n_i - 1} \exp \left( -x/\Omega_i\right) \mathrm dx = \frac{1}{\Gamma(\nu/2) \rho^{n_i}} G_{1, 2}^{2, 1} \left( \frac{1}{\Omega_i \rho} \left\vert \begin{smallmatrix} 1 - n_i \\ 0, \tfrac{\nu}{2} - n_i \end{smallmatrix} \right.  \right)$, $\hat I_{i, 2}(l) \triangleq \int_0^\infty (1 + \rho x)^{-\nu/2} \exp \left( -\Phi_{i, l} x\right) \mathrm dx = \left( \Phi_{i, l}^{(\nu/2) - 1}/\rho^{\nu/2} \right) \exp \left( \Phi_{i, l}/\rho \right) \Gamma \left( 1 - \frac{\nu}{2}, \Phi_{i, l}/\rho \right)$, and $\hat I_{i, 3} (m) \triangleq \int_0^\infty (1 + \rho x)^{-\nu/2} x^m \exp \left( -x/\Omega_i \right) \mathrm dx = \tfrac{1}{\Gamma(\nu/2) \rho^{m + 1}} G_{1, 2}^{2, 1} \left( \tfrac{1}{\Omega_i \rho} \left\vert \begin{smallmatrix} -m \\ 0,\  \tfrac{\nu}{2} - m - 1 \end{smallmatrix} \right.  \right)$. The integrals in $\hat I_{i, 1}$, $\hat I_{i, 2}(l)$ and $\hat I_{i, 3}(m)$ are solved similar to $I_1$, $I_2(l)$ and $I_3(m)$ respectively. The sum effective capacity for OMA-GSC is then obtained by $\hat E_{\mathrm{sum}} = \hat E_w + \hat E_s$.
%-----------------------------------------------------
\subsection{High-SNR approximation for the EC of NOMA-GSC}
In this subsection, we present a high-SNR approximation for the EC of NOMA-GSC. For $\rho \gg 1$, it can be noted from~\eqref{Ei_Def} that, for $x_s$, we have 
\begin{align}
	& \ E_s \approx -\dfrac{1}{\nu} \log_2\left[ \mathbb E_{g_s} \left\{ (a_s \rho g_s)^{-\nu}\right\}\right] \notag \\
	= & \ \log_2 (a_s \rho) - \dfrac{1}{\nu} \log_2 \left[ \int_0^\infty x^{-\nu} f_{g_s}(x) \mathrm dx\right] \notag \\
	= & \ \log_2(a_s \rho) -\dfrac{1}{\nu} \log_2 \left[ \binom{N_s}{n_s} \left( \dfrac{\Gamma(n_s - \nu)}{\Omega_s^{\nu} \Gamma(n_s)} + \dfrac{1}{\Omega_s} \sum_{l = 1}^{N_s - n_s} \right. \right. \notag \\
	& (-1)^{n_s + l - 1} \binom{N_s - n_s}{l} \left( \dfrac{n_s}{l} \right)^{n_s - 1}  \left\{ \Gamma(1-\nu) \Phi_{s, l}^{\nu - 1}  \right. \notag \\
	& \left. \left. \left. - \sum_{m = 0}^{n_s - 2} \dfrac{\Gamma(m - \nu + 1) \Omega_{s}^{m - \nu + 1}}{m!} \left( \dfrac{-l}{n_s \Omega_s}\right)^{m} \right\} \right) \right], \label{Es_High}
\end{align}
where the integral is solved using~\cite[eqn.~(3.351-3),~p.~340]{Grad} and the integral holds good for $\nu < \min\{n_s, 1, m + 1\} = 1$. A similar limitation was encountered in the analysis in~\cite{Larsson}. On the other hand, for the case of $x_w$, a high-SNR approximation for the EC of $x_w$ can be given by 
\begin{align}
	E_w \approx \log_2 \left( 1 + \dfrac{a_w}{a_s} \right). \label{Ew_High}
\end{align}
It is interesting to note from~\eqref{Ew_High} that at high SNR, the effective capacity of $x_w$ is independent of $\theta$ and $N_w$. This means that at high SNR, the EC of $x_w$ becomes (almost) equal to the average achievable rate of $x_w$ and no gain is obtained (in terms of the EC of $x_w$) by having multiple antennas at $U_w$. For $\rho \gg 1$, an approximation for the sum EC can be obtained by adding~\eqref{Es_High} and~\eqref{Ew_High}.
%-----------------------------------------------------
\subsection{Low-SNR approximation for the EC of NOMA-GSC}
A low-SNR ($\rho \to 0$) approximation for the EC can be given by~(c.f.~\cite[eqn.~(18)]{Larsson})
\begin{align}
	E_i = \rho \dot E_i + 0.5 \rho^2 \ddot E_i + O(\rho^2), \label{LowSNRDef}
\end{align}
where $\dot E_i$ and $\ddot E_i$ denote the first and second order derivatives of the EC in~\eqref{Ei_Def} with respect to $\rho$ and evaluated at $\rho = 0$, and $O(\cdot)$ denotes the Landau symbol. Following a similar line of reasoning as in~\cite[\textsc{Appendix}~I]{Derivative}, $\dot E_s$ and $\ddot E_s$ are given by 
\begin{equation}
\begin{aligned}
	& \dot E_s = \log_2(e) a_s \mathbb E(g_s), \\
	& \ddot E_s = \log_2(e) a_s^2 \left[ \nu \{\mathbb E(g_s)\}^2 - (\nu + 1) \mathbb E(g_s^2)\right], \label{Es_dots}
\end{aligned}
\end{equation}
where 
\begin{align*}
	\mathbb E(g_s) = & \ \binom{N_s}{n_s}\left[ n_s \Omega_s + \dfrac{1}{\Omega_s} \sum_{l = 1}^{N_s - n_s} (-1)^{n_s + l - 1} \binom{N_s-n_s}{l}  \right. \\
	& \left. \times \left( \dfrac{n_s}{l} \right)^{n_s - 1} \left\{ \Phi_{s, l}^{-2} - \sum_{m = 0}^{n_s - 2} (m + 1) \Omega_s^2 \left( \dfrac{-l}{n_s}\right)^m\right\} \right],
\end{align*}
\begin{align*}
	\mathbb E (g_s^2) = & \ \binom{N_s}{n_s} \left[ \dfrac{\Gamma(n_s + 2) \Omega_s^{n_s + 2}}{\Omega_s^{n_s} (n_s - 1)!} + \dfrac{1}{\Omega_s} \sum_{l = 1}^{N_s - n_s} (-1)^{n_s + l - 1}  \right. \\
	& \times \binom{N_s - n_s}{l}\left( \dfrac{n_s}{l}\right)^{n_s - 1} \left\{ 2\left( \dfrac{\Omega_s}{\Phi_{s, l}} \right)^3 - \sum_{m = 0}^{n_s - 2} \dfrac{1}{m!} \right. \\
	& \left. \left. \times \left( \dfrac{-l}{n_s \Omega_s}\right)^m \Gamma(m + 3) \Omega_s^{m + 3} \right\} \right].
\end{align*}
The integrals in $\mathbb E(g_s)$ and $\mathbb E(g_s^2)$ are solved using~\cite[eqn.~(3.351-3),~p.~340]{Grad}. Therefore, a closed-form expression for the low-SNR approximation of $E_s$ can be obtained using~\eqref{LowSNRDef} and~\eqref{Es_dots}. Similarly, for $x_w$, we have 
\begin{equation}
\begin{aligned}
	& \dot E_w = \log_2(e) a_w \mathbb E(g_{\min}), \\
	& \ddot E_w = \log_2(e) a_w \left[ \nu a_w \left\{ \mathbb E(g_{\min})\right\}^2 \right. \\
	& \hspace{3cm}\left. - \left\{ (\nu + 1) a_w - 2 a_s \right\} \mathbb E(g_{\min}^2)\right] \label{Ew_dots}
\end{aligned}
\end{equation}
Since it is somewhat tedious to obtain closed-form expressions of $\mathbb E(g_{\min})$ and $\mathbb E(g_{\min}^2)$, we present closed-form analysis for two special cases.
\paragraph*{Case~I ($n_s = n_w = n = 1$)} In this case, NOMA-GSC reduces to NOMA-SC. Therefore, we have 
\begin{align*}
	& \mathbb E(g_{\min}) = \mathbb E(g_{\min, \mathrm{SC}}) = \int_0^\infty x f_{g_{\min, \mathrm{SC}}}(x) \mathrm dx \\
	= & \ \sum_{k = 1}^{N_s} \sum_{j = 1}^{N_w} \binom{N_s}{k} \binom{N_w}{j} \dfrac{(-1)^{k + j}}{\chi_{k, j}}, 
\end{align*}
\begin{align*}
	& \mathbb E(g_{\min}^2) = \mathbb E(g_{\min, \mathrm{SC}}^2) = \int_0^\infty x^2 f_{g_{\min, \mathrm{SC}}}(x) \mathrm dx \\
	= & \ \sum_{k = 1}^{N_s} \sum_{j = 1}^{N_w} \binom{N_s}{k} \binom{N_w}{j} \dfrac{2(-1)^{k + j}}{\chi_{k, j}^2}. 
\end{align*}
Using the two preceding expressions together with~\eqref{LowSNRDef} and~\eqref{Ew_dots}, we can obtain a closed-form expression for the low-SNR approximation of $E_w$ for $n_s = n_w = 1$.
\paragraph*{Case~II: ($n_s = N_s$, $n_w = N_w$)} In this case NOMA-GSC reduces to NOMA-MRC. Therefore, 
\begin{align*}
	& \mathbb E(g_{\min}) = \mathbb E(g_{\min, \mathrm{MRC}}) = \int_0^\infty x f_{g_{\min, \mathrm{MRC}}}(x) \mathrm dx \\
	= & \dfrac{1}{\Gamma(N_s) \Omega_s^{N_s}} \sum_{j = 0}^{N_w - 1} \dfrac{(N_s + j)! \chi_{1, 1}^{-(N_s + j + 1)}}{j! \Omega_w^j} \\
	& \hspace{1.5cm}+ \dfrac{1}{\Gamma(N_w) \Omega_w^{N_w}} \sum_{k = 0}^{N_s - 1} \dfrac{(N_w + k)! \chi_{1, 1}^{-(N_w + k + 1)}}{k! \Omega_s^jk} ,
\end{align*}
\begin{align*}
	& \mathbb E(g_{\min}^2) = \mathbb E(g_{\min, \mathrm{MRC}}^2) = \int_0^\infty x^2 f_{g_{\min, \mathrm{MRC}}}(x) \mathrm dx \\
	= & \ \dfrac{1}{\Gamma(N_s) \Omega_s^{N_s}} \sum_{j = 0}^{N_w - 1} \dfrac{(N_s + j + 1)! \chi_{1, 1}^{-(N_s + j + 2)}}{j! \Omega_w^j} \\
	& \hspace{1.5cm} + \dfrac{1}{\Gamma(N_w) \Omega_w^{N_w}} \sum_{k = 0}^{N_s - 1} \dfrac{(N_w + k + 1)! \chi_{1, 1}^{-(N_w + k + 2)}}{k! \Omega_s^jk}.
\end{align*}
Using the two preceding expressions together with~\eqref{LowSNRDef} and~\eqref{Ew_dots}, we can obtain a closed-form expression for the low-SNR approximation of $E_w$ for $n_s = N_s$ and $n_w = N_w$.
%-----------------------------------------------------
\subsection{Upper-bound on the effective capacity of NOMA-GSC}
In this subsection, we derive a fundamental upper-bound on the EC of NOMA-GSC, using Jensen's inequality. Using the fact that $-\log_2(\cdot)$ is a \emph{log-concave} function, from~\eqref{Ei_Def} it follows that 
\begin{align}
	E_i \leq & \  - \dfrac{1}{\nu} \mathbb E_{\gamma_i} \left[ \log_2 \left\{ (1 + \gamma_i)^{-\nu}\right\} \right] \notag \\
	= & \ \mathbb E_{\gamma_i} \left[ \log_2(1 + \gamma_i)\right] \triangleq \tilde E_i. \label{Ei_tilde}
\end{align}
It is important to note that $\hat E_i$ is independent of the delay exponent $\theta$ and represents the average achievable rate of $x_i$ in NOMA-GSC. his bound is also consistent with the fact that the EC of NOMA-GSC is always less than or equal to the average achievable rate (the equality holds for the case when $\theta \to 0$). A closed-form expression for an upper-bound on $\tilde E_{\mathrm{sum}} = \tilde E_s + \tilde E_w$, which is very tight in the mid-to-high SNR range can be obtained using~\cite[eqns.~(2),~(4)]{EL}.
%==============Results and Discussion========
\section{Results and Discussion}
In this section, we present the numerical and analytical results for the EC of NOMA-GSC and OMA-GSC. We consider a system where $N_s = N_w = N = 4$, $n_s = n_w = n$, $\Omega_s = 1$, $\Omega_w = 0.1$, $T = 0.01$ ms, $B = 100$ kHz. As reported in~\cite{EL}, the optimal power allocation in a NOMA-GSC system depends on the target data rate of $x_i$. Considering the target data rate for both $x_s$ and $x_w$ to be 2 bps/Hz, it follows from~\cite{EL} that a valid range for $a_s$ is given by $0 < a_s < 0.25$. Note that throughout this section, results for the sum EC are presented for the case where the sum EC is optimized using a one-dimensional search over $a_s \in \{ 0.1, 0.2, \ldots, 0.24\}$. Similarly, results for the achievable sum rate are presented for a similarly optimized value of $a_s$.
\begin{figure}[t]
\centering 
\includegraphics[width = 0.9\linewidth]{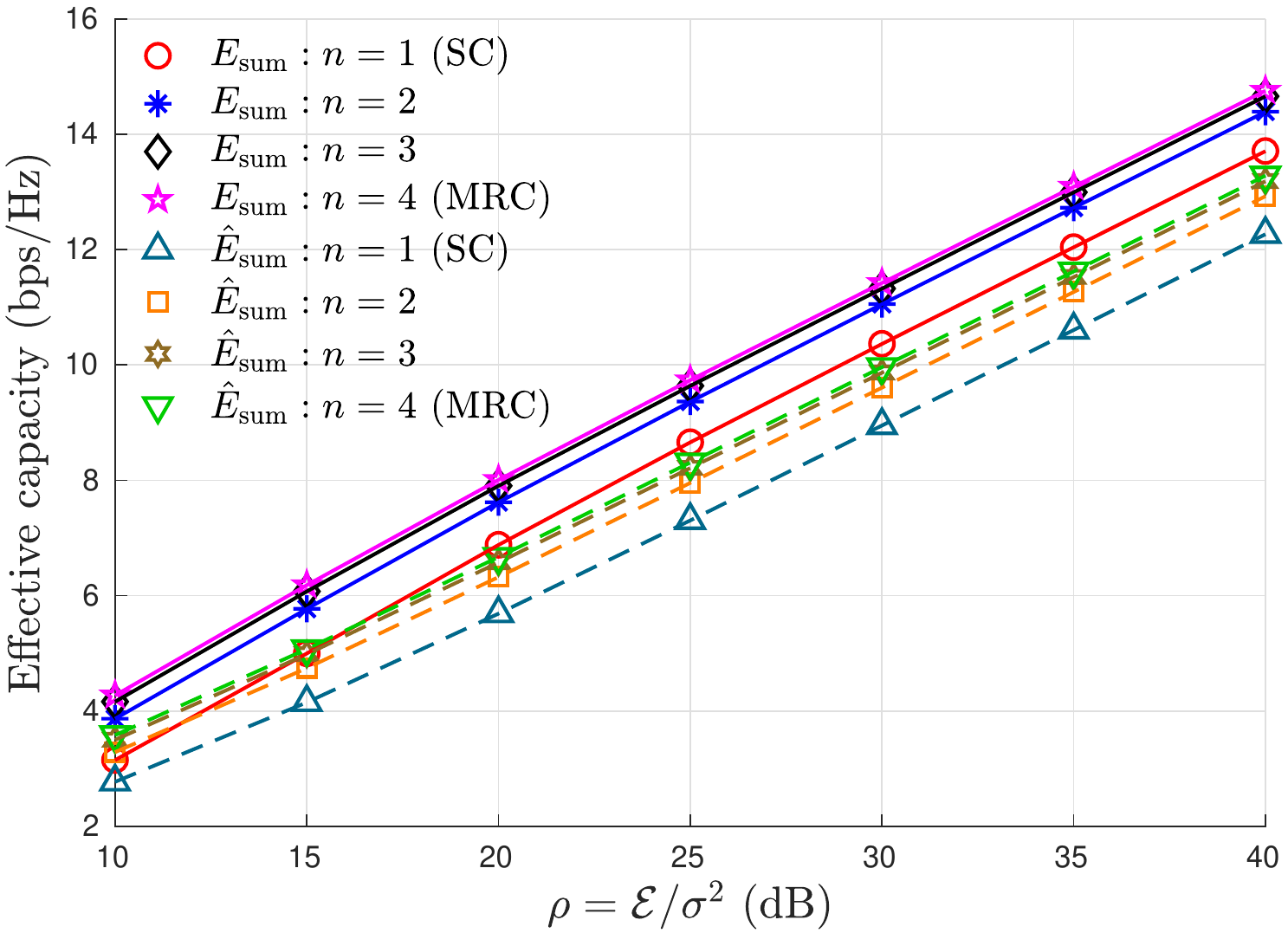}
\caption{Comparison of EC for NOMA-GSC and OMA-GSC with $\theta = 1$.}
\label{NOMA_OMA_Comparison}
\end{figure}
\begin{figure}[t]
\centering 
\includegraphics[width = 0.9\linewidth]{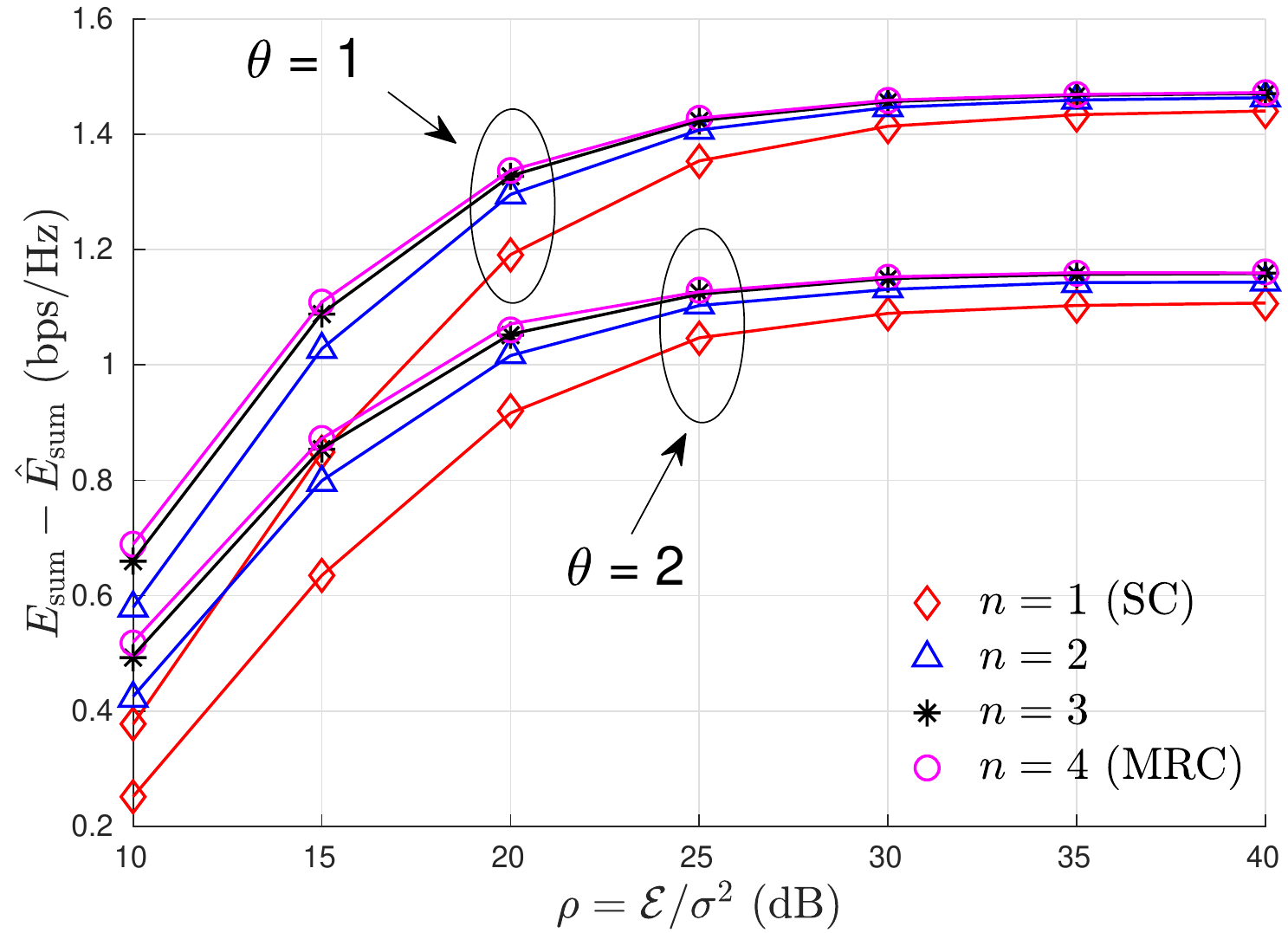}
\caption{Difference between the ECs of NOMA-GSC and OMA-GSC.}
\label{NOMA_OMA_Diff}
\end{figure}
\begin{figure}[t]
\centering 
\includegraphics[width = 0.9\linewidth]{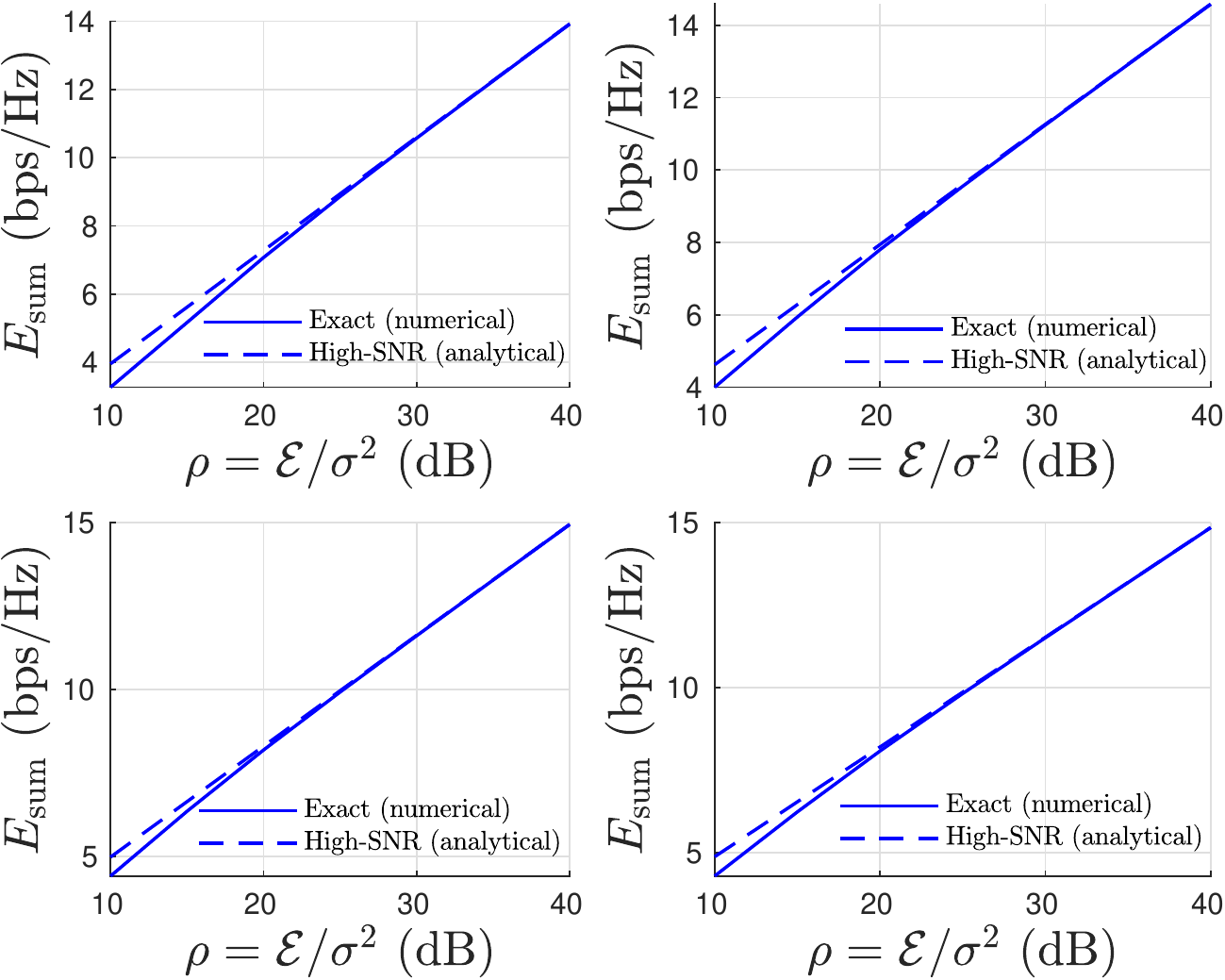}
\caption{High-SNR approximation for the sum EC of NOMA-GSC. Clockwise from the top-left: $n =$ 1 (SC), 2, 3, 4 (MRC).}
\label{HighSNR}
\end{figure}

Fig.~\ref{NOMA_OMA_Comparison} shows a comparison of EC in NOMA-GSC and OMA-GSC for different values of $n$. Markers in the figure denote the numerically evaluated results. The dashed lines (without markers) denote the closed-form analytical results for the sum EC of OMA-GSC. For the case of NOMA-GSC, the solid lines (without markers) for $n = 1$ and $n = 4$ denote the closed-form analytical results, whereas the solid lines (without markers) for $n = 2$ and $n = 3$ represent the semi-analytical results where $E_s$ is evaluated using the closed-form expression and $E_w$ is evaluated numerically. An excellent agreement between the numerical and analytical/semi-analytical results confirms the correctness of the analysis. The optimal value of $a_s$ that maximizes the sum EC of NOMA-GSC is found to be 0.24 (the maximum possible value considered in the range $0 < a_s < 0.24$). This occurs because most of the EC in $E_{\mathrm{sum}}$ is obtained by $E_s$ (as $x_s$ is decoded without any inter-symbol interference and the links between $S$ and $U_s$ are comparatively stronger).

It can be noted from the figure that NOMA-GSC outperforms OMA-GSC for any value of $n$. More interestingly, it can be observed that the gain in the sum EC achieved by moving from $n$ to $n+1$ combined paths decreases with increasing $n$. These results are further elaborated in~Fig.~\ref{NOMA_OMA_Diff}. In the figure, markers denote the numerically evaluated results, whereas the solid lines (without markers) denote the analytical/semi-analytical results. It can be noted from the figure that the difference between $E_{\mathrm{sum}}$ and $\hat E_{\mathrm{sum}}$ increases in the low-to-mid SNR regime, with increasing $\rho$ and also with increasing $n$. This difference then saturates in the high-SNR range (which is in line with the results in~\cite{NOMA_EC_Mussavian}) and most of the gain is obtained by using only two strongest diversity paths. As the value of $\theta$ increases, the figure shows a decrease in the performance difference between NOMA and OMA, which means that the performance of NOMA and OMA both degrade severely under a stringent delay requirement.
\begin{figure}[t]
\centering 
\includegraphics[width = 0.9\linewidth]{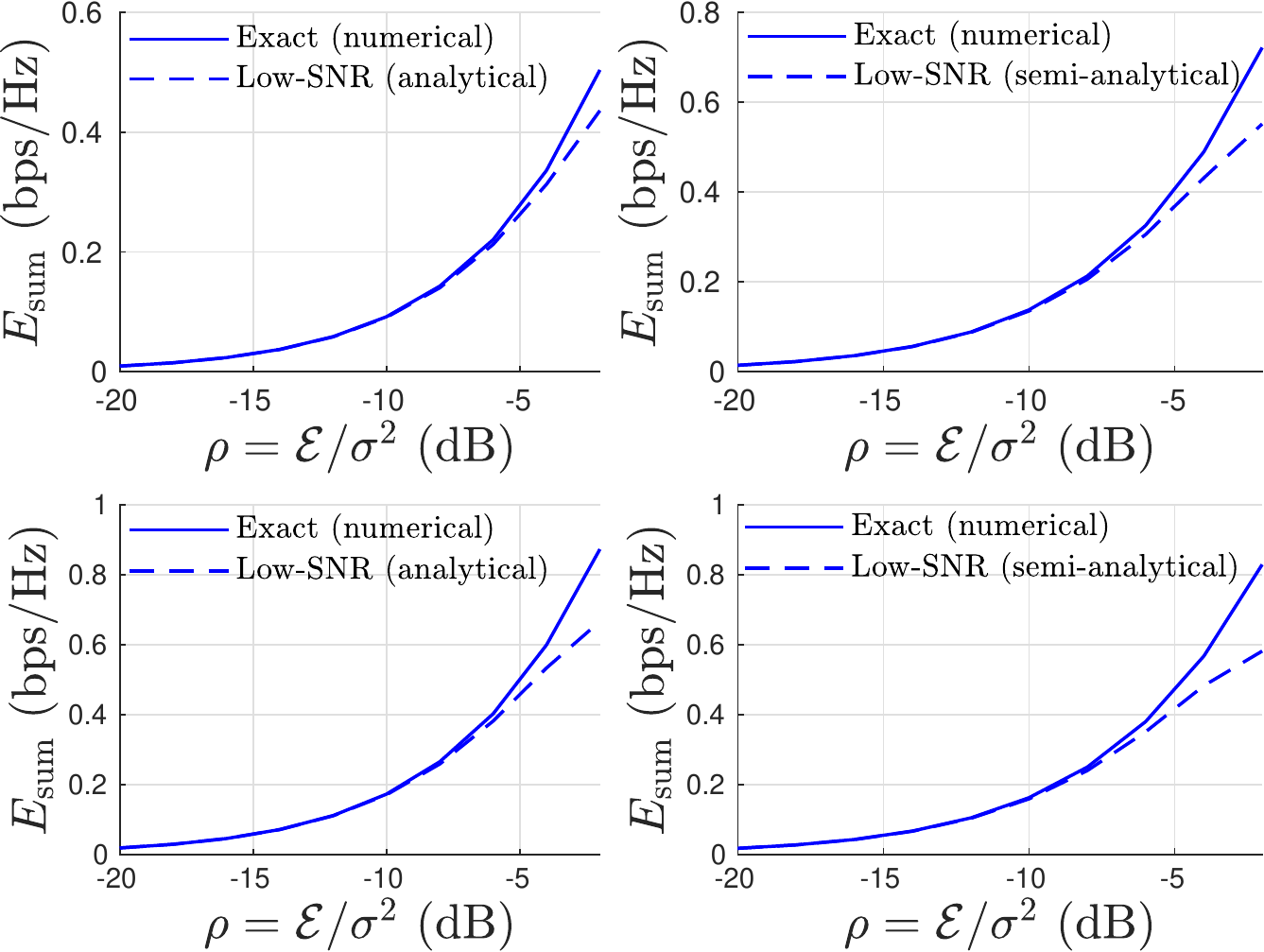}
\caption{Low-SNR approximation for the sum EC of NOMA-GSC. Clockwise from the top-left: $n =$ 1 (SC), 2, 3, 4 (MRC).}
\label{LowSNR}
\end{figure}
\begin{figure}[t]
\centering 
\includegraphics[width = 0.9\linewidth]{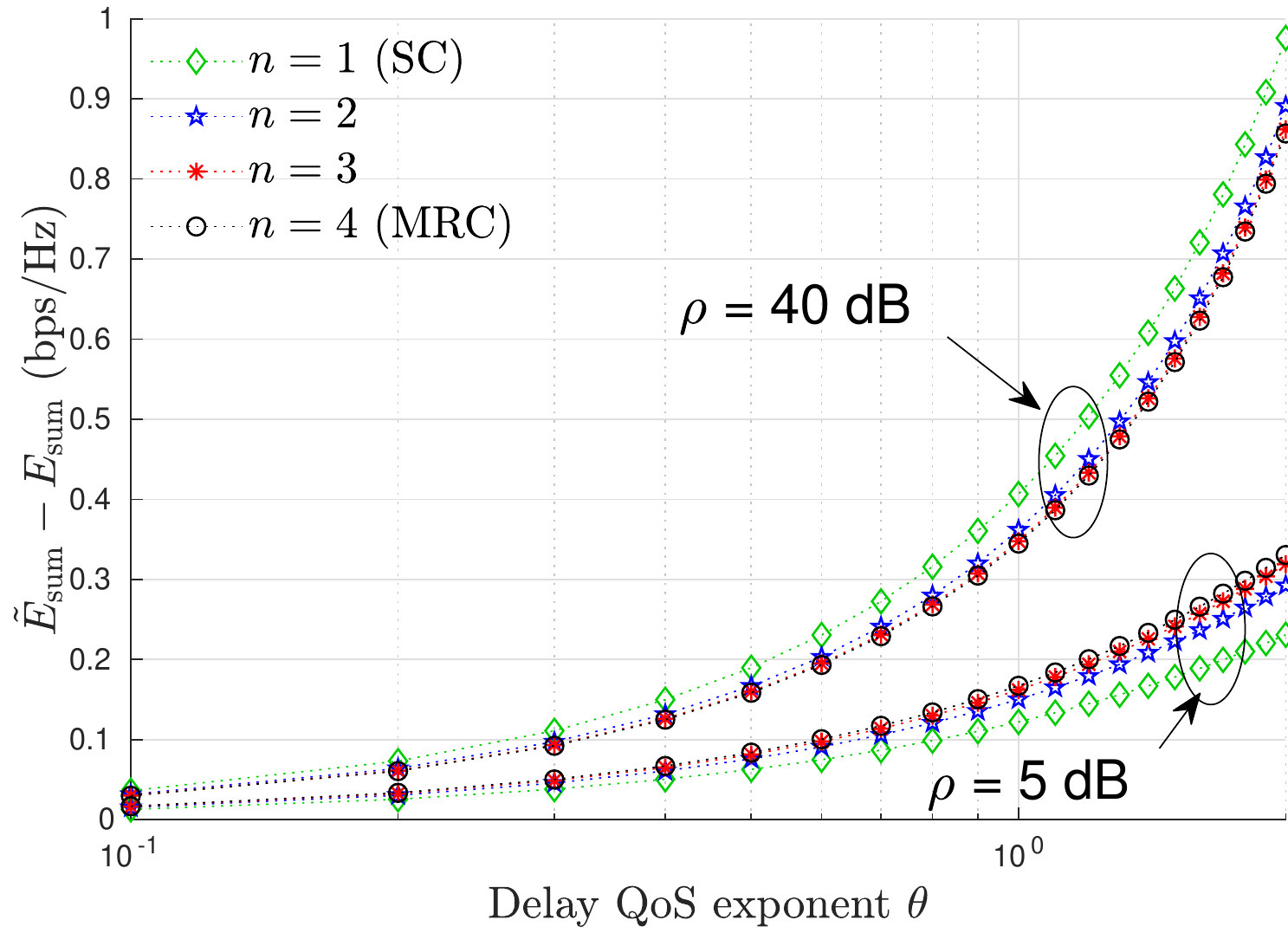}
\caption{Difference between average achievable rate and effective capacity for different values of $\theta$ and $n$.}
\label{RateDiff}
\end{figure}

Figs.~\ref{HighSNR} and~\ref{LowSNR} show the high-SNR and low-SNR approximations for the sum EC of NOMA-GSC for $\theta = 0.5$, respectively. An excellent match can be noticed from both the figures between the exact (numerically evaluated) values of $E_{\mathrm{sum}}$ and the analytically/semi-analytically evaluated approximations. It can be noted from these figures that there is an exponential growth in the sum EC in the low-SNR regime, and a linear growth in the high-SNR regime.

Fig.~\ref{RateDiff} shows the effect of the delay QoS constraint $\theta$ on the difference between the average achievable sum rate and EC of NOMA-GSC (i.e., $\tilde E_{\mathrm{sum}} - E_{\mathrm{sum}}$) for different values of $n$. First, it can be noted from the figure that an increase in the value of $\theta$ results in a very severe system performance degradation as the value of $\tilde E_{\mathrm{sum}} - E_{\mathrm{sum}}$ grows very rapidly for large $\theta$. This means that a system with strict delay constraints will have a much lower EC as compared to the average achievable rate. Furthermore, it can be noted from the figure that with an increase in the value of $\rho$, the difference between the average achievable rate and the effective capacity increases. It is also interesting to note that for low values of $\rho$, although increasing the number of combined paths increases both $\tilde E_{\mathrm{sum}}$ and $E_{\mathrm{sum}}$, the difference between these two quantities also increases. For large values of $\rho$, by combining more diversity paths at the receiver, the difference between $\tilde E_{\mathrm{sum}}$ and $E_{\mathrm{sum}}$ decreases. Therefore, in order to achieve a desired quality of service in terms of delay QoS exponent and data rate, a  higher transmit power requirement is need than that predicted by the traditional achievable rate analysis.
%==============CONCLUSION====================
\section{Conclusion}
In this paper, we have presented the effective capacity analysis of a two-user single-input multiple-output downlink NOMA system with generalized selection combining receivers. We derived closed-form expressions for the sum EC of NOMA-GSC and OMA-GSC. We also presented high and low-SNR approximations for the sum EC of NOMA-GSC. Our results indicate that the sum EC grows exponentially in the low-SNR regime and linearly in the high-SNR regime. The high-SNR analysis of the sum EC confirmed that no benefit is obtained in terms of EC by having multiple antennas at the weak user. The results also indicate that most of the gain in terms of sum EC is obtained by combining the signals received from the strongest diversity paths, while diminishing returns are obtained by increasing the number of combined paths. By quantifying the difference between average achievable sum rate and sum effective capacity, we showed that for a system with stringent delay requirements,
the achievable link-layer rate is significantly smaller than the ergodic rate, and this difference increases with an increase in SNR. The presented analysis serves as a practical system design tool which can be efficiently applied to any configuration in order to determine the appropriate number of diversity paths to be combined to achieve a delay-constrained target quality-of-service.
%==============ACK====================
\balance
%\section*{Acknowledgment}
%This publication has emanated from research conducted with the financial support of Science Foundation Ireland (SFI) and is co-funded under the European Regional Development Fund under Grant Number 13/RC/2077.

%%=============References=============
\Urlmuskip=0mu plus 1mu\relax
\bibliographystyle{IEEEtran}
\bibliography{ICC2020}

% Generated by IEEEtran.bst, version: 1.14 (2015/08/26)
\begin{thebibliography}{10}
\providecommand{\url}[1]{#1}
\csname url@samestyle\endcsname
\providecommand{\newblock}{\relax}
\providecommand{\bibinfo}[2]{#2}
\providecommand{\BIBentrySTDinterwordspacing}{\spaceskip=0pt\relax}
\providecommand{\BIBentryALTinterwordstretchfactor}{4}
\providecommand{\BIBentryALTinterwordspacing}{\spaceskip=\fontdimen2\font plus
\BIBentryALTinterwordstretchfactor\fontdimen3\font minus
  \fontdimen4\font\relax}
\providecommand{\BIBforeignlanguage}[2]{{%
\expandafter\ifx\csname l@#1\endcsname\relax
\typeout{** WARNING: IEEEtran.bst: No hyphenation pattern has been}%
\typeout{** loaded for the language `#1'. Using the pattern for}%
\typeout{** the default language instead.}%
\else
\language=\csname l@#1\endcsname
\fi
#2}}
\providecommand{\BIBdecl}{\relax}
\BIBdecl

\bibitem{ProcHanzo}
Y.~{Liu}, Z.~{Qin}, M.~{Elkashlan}, Z.~{Ding}, A.~{Nallanathan}, and
  L.~{Hanzo}, ``Nonorthogonal multiple access for {5G} and beyond,''
  \emph{Proc. of the IEEE}, vol. 105, no.~12, pp. 2347--2381, Dec 2017.

\bibitem{NOMA_book}
M.~Vaezi, Z.~Ding, and H.~Poor, \emph{Multiple Access Techniques for {5G}
  Wireless Networks and Beyond}.\hskip 1em plus 0.5em minus 0.4em\relax
  Springer International Publishsing, 2018.

\bibitem{Coop_NOMA}
Z.~{Ding}, M.~{Peng}, and H.~V. {Poor}, ``Cooperative non-orthogonal multiple
  access in {5G} systems,'' \emph{IEEE Commun. Lett.}, vol.~19, no.~8, pp.
  1462--1465, Aug 2015.

\bibitem{MenGe}
J.~{Men} and J.~{Ge}, ``Non-orthogonal multiple access for multiple-antenna
  relaying networks,'' \emph{IEEE Commun. Lett.}, vol.~19, no.~10, pp.
  1686--1689, Oct 2015.

\bibitem{MIMO-NOMA}
M.~{Zeng}, A.~{Yadav}, O.~A. {Dobre}, G.~I. {Tsiropoulos}, and H.~V. {Poor},
  ``On the sum rate of {MIMO-NOMA} and {MIMO-OMA} systems,'' \emph{IEEE
  Wireless Commun. Lett.}, vol.~6, no.~4, pp. 534--537, Aug 2017.

\bibitem{Negi}
{Dapeng Wu} and R.~{Negi}, ``Effective capacity: a wireless link model for
  support of quality of service,'' \emph{IEEE Trans. Wireless Commun.}, vol.~2,
  no.~4, pp. 630--643, July 2003.

\bibitem{Larsson}
M.~{Matthaiou}, G.~C. {Alexandropoulos}, H.~Q. {Ngo}, and E.~G. {Larsson},
  ``Analytic framework for the effective rate of {MISO} fading channels,''
  \emph{IEEE Trans. Commun.}, vol.~60, no.~6, pp. 1741--1751, June 2012.

\bibitem{Snedecor}
S.~{Chen}, J.~{Zhang}, G.~K. {Karagiannidis}, and B.~{Ai}, ``Effective rate of
  {MISO} systems over {Fisher}-{Snedecor} $\mathcal{F}$ fading channels,''
  \emph{IEEE Commun. Lett.}, vol.~22, no.~12, pp. 2619--2622, Dec 2018.

\bibitem{NOMA_EC_Mussavian}
W.~{Yu}, L.~{Musavian}, and Q.~{Ni}, ``Link-layer capacity of {NOMA} under
  statistical delay {QoS} guarantees,'' \emph{IEEE Trans. Commun.}, vol.~66,
  no.~10, pp. 4907--4922, Oct 2018.

\bibitem{SubOptimal}
J.~{Choi}, ``Effective capacity of {NOMA} and a suboptimal power control policy
  with delay {QoS},'' \emph{IEEE Trans. Commun.}, vol.~65, no.~4, pp.
  1849--1858, April 2017.

\bibitem{OptimalPower}
X.~{Chen}, G.~{Liu}, and Z.~{Ma}, ``Statistical {QoS} provisioning for
  half/full-duplex cooperative non-orthogonal multiple access,'' in \emph{2017
  IEEE 86th Veh. Tech. Conf. (VTC-Fall)}, Sep. 2017, pp. 1--5.

\bibitem{NOMA_JSTSP}
C.~{Xiao}, J.~{Zeng}, W.~{Ni}, R.~P. {Liu}, X.~{Su}, and J.~{Wang}, ``Delay
  guarantee and effective capacity of downlink {NOMA} fading channels,''
  \emph{IEEE J. Sel. Topics Signal Proc.}, vol.~13, no.~3, pp. 508--523, June
  2019.

\bibitem{NOMA_Secrecy}
W.~{Yu}, A.~{Chorti}, L.~{Musavian}, H.~V. {Poor}, and Q.~{Ni}, ``Effective
  secrecy rate for a downlink {NOMA} network,'' \emph{IEEE Trans. Wireless
  Commun.}, to appear.

\bibitem{MGF-GSC}
M.~S. {Alouini} and M.~K. {Simon}, ``An {MGF}-based performance analysis of
  generalized selection combining over {R}ayleigh fading channels,'' \emph{IEEE
  Trans. Commun.}, vol.~48, no.~3, pp. 401--415, March 2000.

\bibitem{EL}
\BIBentryALTinterwordspacing
V.~{Kumar}, B.~{Cardiff}, and M.~F. {Flanagan}, ``Performance analysis of
  {NOMA} with generalised selection combining receivers,'' \emph{Electronics
  Letters}, to appear. [Online]. Available:
  \url{https://www.researchgate.net/publication/336326352_Performance_analysis_of_NOMA_with_generalized_selection_combining_receiverss}
\BIBentrySTDinterwordspacing

\bibitem{Reduce}
V.~S. Adamchik and O.~I. Marichev, ``The algorithm for calculating integrals of
  hypergeometric type functions and its realization in {REDUCE} system,'' in
  \emph{Proceedings of the International Symposium on Symbolic and Algebraic
  Computation}, 1990, pp. 212--224.

\bibitem{Grad}
A.~Jeffrey and D.~Zwillinger, \emph{Table of Integrals, Series, and Products},
  7th~ed.\hskip 1em plus 0.5em minus 0.4em\relax Elsevier Science, 2007.

\bibitem{KappaMuShadowed}
C.~Garc\'ia-Corrales, F.~J. Ca{\~n}ete, and J.~F. Paris, ``Capacity of
  $\kappa-\mu$ shadowed fading channels,'' \emph{International Journal of
  Antennas and Propagation}, vol.~24, 2014.

\bibitem{Derivative}
C.~{Zhong}, T.~{Ratnarajah}, K.~{Wong}, and M.~{Alouini}, ``Effective capacity
  of correlated {MISO} channels,'' in \emph{2011 IEEE Int.Conf. Commun. (ICC)},
  June 2011, pp. 1--5.

\end{thebibliography}

\end{document}